\documentclass[prb,aps,superscriptaddress,reprint]{revtex4-1}

\usepackage{graphicx}
\usepackage{dcolumn}
\usepackage{bm}
\usepackage{setspace}
\usepackage{amsmath}
\usepackage{epsfig}

\begin{document}

\title{Understanding Electrical Conduction and Nanopore Formation During Controlled Breakdown}

\author{Jasper P.~Fried}
\author{Jacob L.~Swett}
\affiliation{Department of Materials, University of Oxford, Oxford, OX1 3PH, U.K.}
\author{Binoy Paulose Nadappuram}
\author{Aleksandra Fedosyuk}
\affiliation{Department of Chemistry, Imperial College London, Molecular Sciences Research Hub, White City Campus, 82 Wood Lane, W12 0BZ, U.K.}
\author{Pedro Miguel Sousa}
\affiliation{Instituto de Tecnologia Química e Biológica António Xavier, Universidade Nova de Lisboa, Av. da República, 2780-157 Oeiras, Portugal}
\author{Dayrl P. Briggs}
\affiliation{Center for Nanophase Materials Sciences, Oak Ridge National Laboratory, Oak Ridge, TN 37830, U.S.A.}
\author{Aleksandar P. Ivanov}
\author{Joshua B. Edel}
\affiliation{Department of Chemistry, Imperial College London, Molecular Sciences Research Hub, White City Campus, 82 Wood Lane, W12 0BZ, U.K.}
\author{Jan A.~Mol}
\affiliation{School of Physics and Astronomy, Queen Mary University of London, London, U.K.}
\author{James R.~Yates}
\affiliation{Instituto de Tecnologia Química e Biológica António Xavier, Universidade Nova de Lisboa, Av. da República, 2780-157 Oeiras, Portugal}

\date{\today}



\begin{abstract}

Controlled breakdown has recently emerged as a highly appealing technique to fabricate solid-state nanopores for a wide range of biosensing applications. This technique relies on applying an electric field of approximately 0.6-1$\,$V/nm across the membrane to induce a current, and eventually, breakdown of the dielectric. However, a detailed description of how electrical conduction through the dielectric occurs during controlled breakdown has not yet been reported. Here, we study electrical conduction and nanopore formation in SiN$_x$ membranes during controlled breakdown. We show that depending on the membrane stoichiometry, electrical conduction is limited by either oxidation reactions that must occur at the membrane-electrolyte interface (Si-rich SiN$_x$), or electron transport across the dielectric (stoichiometric SiN$_x$). We provide several important implications resulting from understanding this process which will aid in further developing controlled breakdown in the coming years, particularly for extending this technique to integrate nanopores with on-chip nanostructures. 

\end{abstract}

\maketitle

Nanopore sensors consist of a nanometre sized hole in an insulating membrane that separates two chambers of electrolyte solution. When a voltage is applied across the membrane, ions flow through the nanopore resulting in a measurable ionic current. When a biomolecule is drawn into and through the nanopore, it affects the passage of ions resulting in a change in the ionic current. Measuring such changes in the ionic current therefore provides a simple single-molecule biosensing technique \cite{Xue2020,Miles2013,Varongchayakul2018,Albrecht2019}. Indeed, over the past several decades, nanopores have proven to be versatile single-molecule biosensing devices with applications ranging from DNA \cite{Manrao2012,Feng2015} and protein sequencing \cite{Restrepo-Perez2018,Ouldali2019}, to ultra-dilute analyte detection \cite{Freedman2016,Rozevsky2020,Chuah2019,Wu2020}, polymer data storage \cite{Bell2016,Chen2020}, and enzymology \cite{Willems2017}.

Nanopore sensors can be classified as either biological \cite{Howorka2017} or solid-state \cite{Dekker2007}. Biological nanopores generally consist of barrel shaped proteins that self-insert into lipid or synthetic membranes. Solid-state nanopores, however, are typically formed in thin ($<$50$\,$nm) dielectrics such as SiN$_x$ \cite{Li2001}, TiO$_2$, \cite{Wang2018a} and HfO$_2$ \cite{Larkin2013} or two-dimensional materials such as graphene \cite{Merchant2010,Garaj2010,Schneider2010}, MoS$_2$ \cite{Feng2015}, and hBN \cite{Liu2013}. The ability to fabricate solid-state nanopores of different diameters and operate them in a wide range of environmental conditions makes them particularly attractive for many of the applications discussed above \cite{Dekker2007,Xue2020}. In the past, solid-state nanopores were typically fabricated using focused charged particle beams to locally sputter material from the membrane \cite{Kim2006,Wu2009,Gierak2007,Lo2006}. However, this requires specialised equipment, trained operators, and is a labour intensive process thus limiting the availability of this technique to the wider research community.

\begin{figure*}
    \centering
    \includegraphics[width=\textwidth]{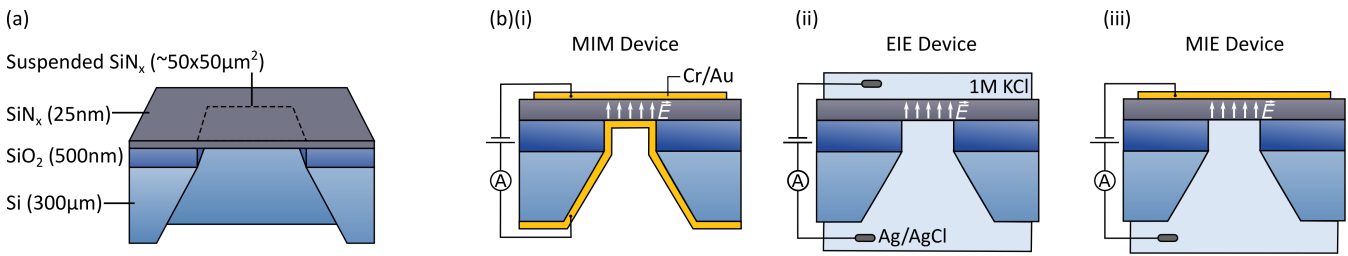}
    \caption{(a) Schematic of the basic device geometry used in this work (note that SiO$_2$ and SiN$_x$ layers are not shown on the backside of the device for simplicity). (b) Schematics of the experimental setup used for (i) metal-insulator-metal (MIM), (ii) electrolyte-insulator-electrolyte (EIE), and (iii) metal-insulator-electrolyte (MIE) devices.}
    \label{fdevices}
\end{figure*}

To overcome these issues, a technique called controlled breakdown (CBD) has been developed to fabricate nanopores in solid-state membranes \cite{Kwok2014,Waugh2020,Fried2021}. In this method, an electric field of approximately 0.4-1$\,$V/nm is applied across the membrane via the electrolyte solutions whilst simultaneously measuring the resulting current. After a given period, a spike in the current is observed signifying pore formation at which point the voltage is quickly reduced to ensure the fabrication of a small nanopore. This technique has been used to create pores with diameters down to a single nanometre \cite{Waugh2020} in a range of materials \cite{Kwok2014,Kwok2014a,Wang2018a,Wang2017}. The main advantage of CBD is that it does not require highly specialised equipment and can be fully automated \cite{Waugh2020}, thus resulting in a low fabrication cost and time while also removing the need for experienced operators. The accessibility of this method has resulted in CBD becoming a popular solid-state nanopore fabrication technique in recent years \cite{Goto2018,Rozevsky2020,Briggs2017,Bandara2020,Spitzberg2020}.

Despite CBD being used in many studies, the mechanism by which nanopores are formed during this process remains largely unexplored. Nanopore formation is generally assumed to proceed via a similar mechanism to dielectric breakdown in metal-insulator and semiconductor microelectronic devices \cite{Briggs2015}. In particular, electric fields on the order of (0.1-1$\,$V/nm) activate electron transport through charge traps in the dielectric. At some point, these charge traps form a percolation path which results in an abrupt increase in the current and damage to the dielectric, likely due to Joule heating \cite{Briggs2015}. However, for the case of CBD where the electric field is applied via electrolyte solutions the process is more complex. In this case, oxidation/reduction reactions must occur at the membrane-electrolyte interface to inject/remove electrons from the dielectric. The importance of such redox reactions has been raised in previous studies where it was shown that the pH of the electrolyte solution affects the voltage at which breakdown occurs \cite{Kwok2014,Briggs2015,Yanagi2019,Yanagi2020}. Gas formation at the membrane interface resulting from redox reactions has also been observed during CBD \cite{Dong2020}. However, to date, a detailed description of electrical conduction across a dielectric during CBD has not been provided. Better understanding this process will no doubt aid in continuing the development of CBD as a nanopore fabrication technique, e.g. to fabricate nanopores integrated with on-chip nanostructures or in previously unexplored membrane materials.

Here we study the mechanism of electrical conduction and nanopore formation in SiN$_x$ membranes during CBD. We compare conduction and breakdown in SiN$_x$ membranes when the voltage is applied via (i) metal electrodes on the membrane surface, (ii) electrolyte solutions, and (iii) a combination of the two. For Si-rich SiN$_x$ membranes we show that oxidation reactions at the membrane-electrolyte interface limit the conduction across the membrane thereby increasing the voltage required to cause breakdown. One result of this is that when performing CBD on devices with metal electrodes on the membrane surface we can remove the need for oxidation reactions (since electrons can be supplied by the metal) allowing us to localise pore formation to the electrodes. Interestingly, oxidation reactions at the membrane-electrolyte interface no longer limit the conduction for stoichiometric Si$_3$N$_4$ films. Here, the electrical conduction is predominately limited by electron transport across the dielectric which is significantly reduced compared to Si-rich SiN$_x$ thus highlighting the material dependent nature of the CBD process. 

\section{Results and discussion}

A schematic of our device geometry is shown in Fig.~\ref{fdevices}(a). These devices consist of a SiN$_x$ membrane suspended on 500$\,$nm of SiO$_2$ on a 300$\,\mu$m thick Si substrate. The SiO$_2$ layer is typically used in solid-state nanopore devices to reduce the device capacitance and therefore the high frequency noise \cite{Tabard-Cossa2007,Chien2019,Rosenstein2012}. For our experiments, the SiO$_2$ layer has the additional advantage that it ensures the leakage current is only through the suspended region of the SiN$_x$ membrane. Without the SiO$_2$ layer, charge could be transported from the electrolyte solution, to the Si substrate, and to the SiN$_x$ layer \cite{Briggs2015,Yanagi2020}. Unless stated otherwise, results were obtained for a 25$\,$nm thick Si-rich SiN$_x$ membrane with a nitrogen to silicon ratio of N:Si$\equiv$$x$$\equiv$1.14. The stoichiometry was estimated based on the refractive index of the film ($n$=2.14) \cite{Gilboa2019}. The membrane thickness was estimated from ellipsometry measurements. A description of all the wafers used in this study is provided in SI 1. Details of the fabrication process are provided in the Methods section.

Electrical conduction and breakdown was studied in devices when the electric field is applied in three different ways [Fig.~\ref{fdevices}(b)]. Firstly, we apply a voltage via metal electrodes (5/45$\,$nm Cr/Au) deposited on both sides of the membrane [Fig.~\ref{fdevices}(b)(i)]. Similar device geometries have been studied by the microelectronics community for several decades \cite{Sze1967}. We refer to these devices as metal-insulator-metal (MIM). Next, we study the case when a voltage is applied via electrolyte solutions (1$\,$M KCl with 10$\,$mM Tris and 0.1$\,$mM EDTA at pH 8) on either side of the membrane using Ag/AgCl electrodes [Fig.~\ref{fdevices}(b)(ii)]. This is the typical measurement setup for CBD \cite{Kwok2014,Briggs2014,Waugh2020}. We will refer to this device geometry as electrolyte-insulator-electrolyte (EIE). Lastly, we study the case where the voltage is applied between a metal electrode on the membrane surface and an electrolyte solution on the other side of the membrane [Fig.~\ref{fdevices}(b)(iii)]. We refer to this device geometry as metal-insulator-electrolyte (MIE). For each device geometry we study conduction and dielectric breakdown in the SiN$_x$ membrane by applying a voltage ramp (increasing in steps of 100$\,$mV every 4$\,$s) and measuring the resulting current. Devices from the same wafer are used when comparing between these three geometries to reduce variability resulting from the fabrication process. Namely, electrodes were deposited on a subset of devices from a given wafer to make MIM and MIE devices. 

\subsection{Electron Transport through Silicon Nitride}

It is first useful to discuss the case of conduction across SiN$_x$ when the electric field is applied between metal electrode layers on either side of the membrane surface [Fig.~\ref{fdevices}(b)(i)]. For such a device geometry, the measured current is limited by electron transport processes across the dielectric. Such electron transport processes across thin SiN$_x$ films have been well studied and are often attributed to Poole-Frenkel (PF) emission \cite{Habermehl2002,Habermehl2005,Habermehl2009,Sze1967}. This electron transport process results from lowering of the barrier height between trapped electrons and the conduction band when applying electric fields across the dielectric. Lowering of the barrier height increases the probability of trapped electrons being thermally excited to the conduction band, where they briefly transit the membrane before returning to a localised state. The current density resulting from PF emission can be calculated as \cite{Sze1967}:

\begin{equation}
    J(E,T) = C_{1}Ee^{-q(\Phi_{\mathrm{B}}-\sqrt{qE/\pi\epsilon_{\mathrm{D}}})/k_{\mathrm{B}} T}
    \label{epf}
\end{equation}

where $q$ is the electron charge, $k_{\mathrm{B}}$ is Boltzmann's constant, $T$ is the temperature in Kelvin, $\epsilon_{\mathrm{D}}$ is the optical (dynamic) dielectric constant, $\Phi_{\mathrm{B}}$ is the charge trap depth, and $C_1$ is a constant that is determined by the charge trap density and the carrier mobility. The optical dielectric constant can be calculated as $\epsilon_{\mathrm{D}} \sim n^2$ where $n$ is the refractive index of the film \cite{Jeong2005}. Following Eq.~\ref{epf}, if PF emission is the dominant conduction mechanism a plot of $\mathrm{ln}(J/E)$ vs $E^{1/2}$ should be linear. Such a plot is commonly used to study conduction in dielectric films and is referred to as a PF plot. 

\begin{figure}
    \centering
    \includegraphics[width=\columnwidth]{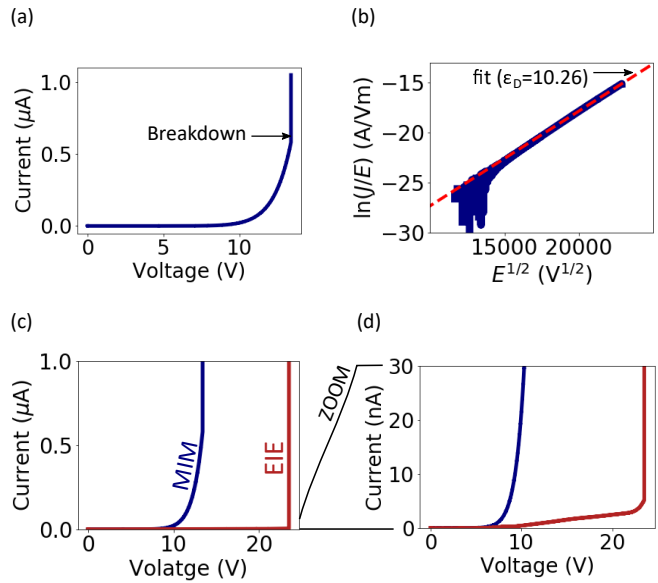}
    \caption{(a) Measured current as a function of voltage for a MIM device. (b) Shows the same data as (a) plotted as a Poole-Frenkel (PF) plot. (c) Measured current as a function of voltage for a EIE device (maroon curve). Also shown in is the measured current as a function of voltage in a MIM device. (d) Shows the same data as (c) with a reduced $y$-scale to enable visualisation of the pre-breakdown conduction behaviour in the EIE device.}    
    \label{fmim}
\end{figure}

Figure \ref{fmim}(a) shows a typical plot of current as a function of applied voltage for a MIM device. We observe an exponential increase in the current before an abrupt spike that indicates dielectric breakdown of the SiN$_x$. Figure \ref{fmim}(b) shows the same data plotted as a PF plot. The PF plot shows a linear behaviour. We also observe an increase in conduction upon increasing the membrane temperature consistent with Eq.~\ref{epf} (SI 2). These results are consistent with PF emission as the dominant electron transport processes (e.g.~rather than direct tunnelling processes which would show different conduction behaviour and no temperature dependence). Note, however, that these results do not guarantee that conduction can be explained solely as a result of PF conduction as given by Eq.~\ref{epf}. For this to be confirmed, it is necessary to extract $\epsilon_{\mathrm{D}}$ from the slope of the PF plot and confirm this is comparable to the value expected from the refractive index of the film \cite{ODwyer1966,Sze1967,Schroeder2015}. We have extracted $\epsilon_{\mathrm{D}}$ by fitting Eq.~\ref{epf} to the data in Fig.~\ref{fmim}(b) (red dashed line) and obtained a value of 10.26 which is significantly higher than the expected value ($\epsilon_{\mathrm{D}}\approx n^2=$ 4.57).

Previous studies on conduction through dielectric films have often demonstrated that despite showing PF like behaviour, the value of $\epsilon_{\mathrm{D}}$ extracted from fitting Eq.~\ref{epf} to the measured data does not match the expected value \cite{Andrews1980,Svensson1977,Pulfrey1970,Schroeder2015}. In many of these studies, it was demonstrated that space charge effects resulting from the trapping of injected charge can significantly affect the conduction behaviour \cite{Andrews1980,Svensson1977,Pulfrey1970}. Such trapping results in a non-uniform charge distribution across the membrane which modifies the electric field. This generally results in slow changes in the measured current as a function of time as the trapped charges accumulate in the dielectric. Consistent with this, we observe slow changes in the current (much slower than those expected from the device capacitance) upon changing the electric field (SI 3). Electron transport in SiN$_x$ films is clearly a complex phenomena that is determined by several processes as well as the specific properties of the film being studied (e.g.~thickness and stoichiometry). A detailed study of this is beyond the scope of this work, however, the above results indicate that PF emission and space charge effects play an important role in determining the conduction in our SiN$_x$ membranes. 

\subsection{Electron transfer reactions at the Electrolyte Membrane interface}

We will now discuss conduction and breakdown in SiN$_x$ membranes when the electric field is applied via electrolyte solutions on each side of the membrane [Fig.~\ref{fdevices}(b)(ii)]. For this device geometry, in addition to electron transport across the dielectric, electron transfer (redox) reactions must also occur at the membrane-electrolyte interface for a current to flow. Previous studies have postulated that the oxidation of Cl$^{-}$ and OH$^{-}$ and the reduction of H$^{+}$ are the dominant redox reactions that occur at the membrane interface when performing CBD in aqueous KCl solutions \cite{Yanagi2020,Dong2020}.

Figure \ref{fmim}(c) shows a typical measurement for electrical conduction and dielectric breakdown in an EIE device (i.e.~nanopore fabrication via CBD). The current measured through a MIM device [as in Fig.~\ref{fmim}(a)] is also shown for comparison. To enable visualisation of the conduction prior to breakdown in the EIE device, Fig.~\ref{fmim}(d), shows the same data as Fig.~\ref{fmim}(c) with a reduced $y$-scale. From these plots, it is clear there is a significant reduction in the measured current for the EIE device. Moreover, a larger voltage must be applied to induce breakdown in the EIE device. These results highlight the importance of redox reactions that must occur at the membrane-electrolyte interface for current to flow in the EIE device. In particular, these redox reactions limit the amount of current transported across the membrane resulting in a larger voltage being required to induce breakdown. Previous studies have pointed out that such redox reactions must be present for a current to flow during CBD \cite{Kwok2014,Yanagi2020,Dong2020}. However, until now it has not been demonstrated that these reactions are the limiting process for conduction during CBD.

\begin{figure}
    \centering
    \includegraphics[width=\columnwidth]{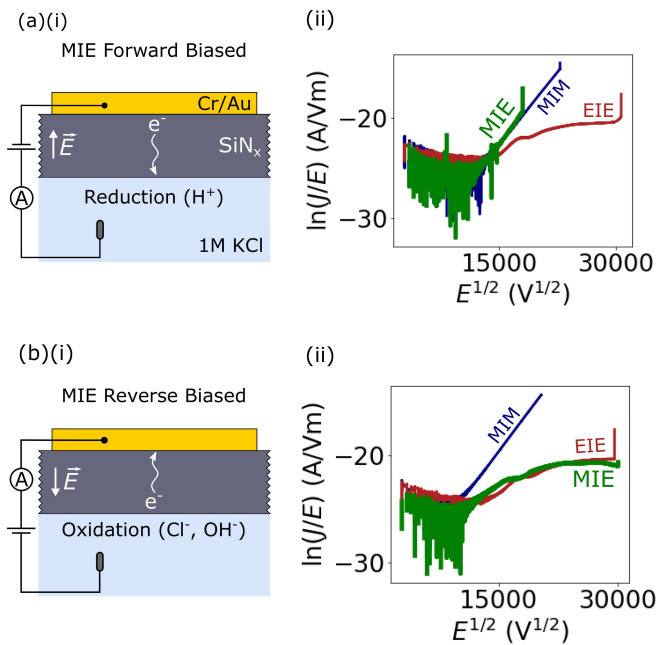}
    \caption{(a)(i) Schematic of the device geometry for a MIE device in the forward-biased electric field configuration. (a)(ii) PF plot of the conduction in a MIM, EIE, and MIE device for the forward-biased configuration. (b)(i) Schematic of the device geometry for a MIE device in the reverse-biased electric field configuration. (b)(ii) PF Plot of the conduction in a MIM, EIE, and MIE device for the reverse-biased configuration.}
    \label{fmie}
\end{figure}

To better understand these redox reactions we have measured conduction and breakdown in devices when the electric field is applied between a metal electrode on one side of the membrane and an electrolyte solution on the other side [MIE devices in Fig.~\ref{fdevices}(b)(iii)]. The asymmetry of this device geometry allows us to isolate contributions from oxidation and reduction reactions occurring on either side of the membrane by changing the direction of the applied field. Grounding the metal electrode and applying a positive voltage to the electrolyte solution results in the electric field direction shown in Fig.~\ref{fmie}(a)(i). We will refer to this as the forward-biased configuration. Reversing the electric field direction by applying a positive voltage to the metal electrode and grounding the electrolyte solution will be referred to as the reverse-biased configuration [Fig.~\ref{fmie}(b)(i)]. 

Figure \ref{fmie}(a)(ii) shows a PF plot comparing conduction in MIM, EIE, and MIE devices for the forward-biased configuration. Conduction in the MIE device shows a similar behaviour to the MIM device with a linear trend on the PF plot. However, when the direction of the electric field is reversed we observe the opposite behaviour [Fig.~\ref{fmie}(b)(ii)]. In particular, conduction in the MIE device now follows a similar behaviour to the EIE device showing a distinctly non-linear trend on the PF plot. As shown in SI 4, this behaviour is reproducible across many devices. The conduction behaviour of the MIM and EIE devices do not change significantly depending on the direction of the applied electric field given the symmetry of these devices.

The change in conduction behaviour of the MIE device upon reversing the direction of the applied electric field provides insight into which redox reaction limits the conduction. For the forward-biased case, an oxidation reaction does not need to occur as electrons can be injected into the SiN$_x$ from the metal electrode. However, a reduction reaction must still occur to remove electrons from the membrane [Fig.~\ref{fmie}(a)(i)]. This configuration results in relatively large conduction through the membrane. For the reverse-biased case, an oxidation reaction must occur to inject electrons into the SiN$_x$ from the electrolyte solution. However, a reduction reaction does not need to occur as electrons can be removed through the metal electrode [Fig.~\ref{fmie}(b)(i)]. This configuration results in a reduced conduction through the membrane. As such, we conclude that oxidation reactions at the membrane-electrolyte interface limit the conduction across the membrane. 

\begin{figure*}
    \centering
    \includegraphics[width=15cm]{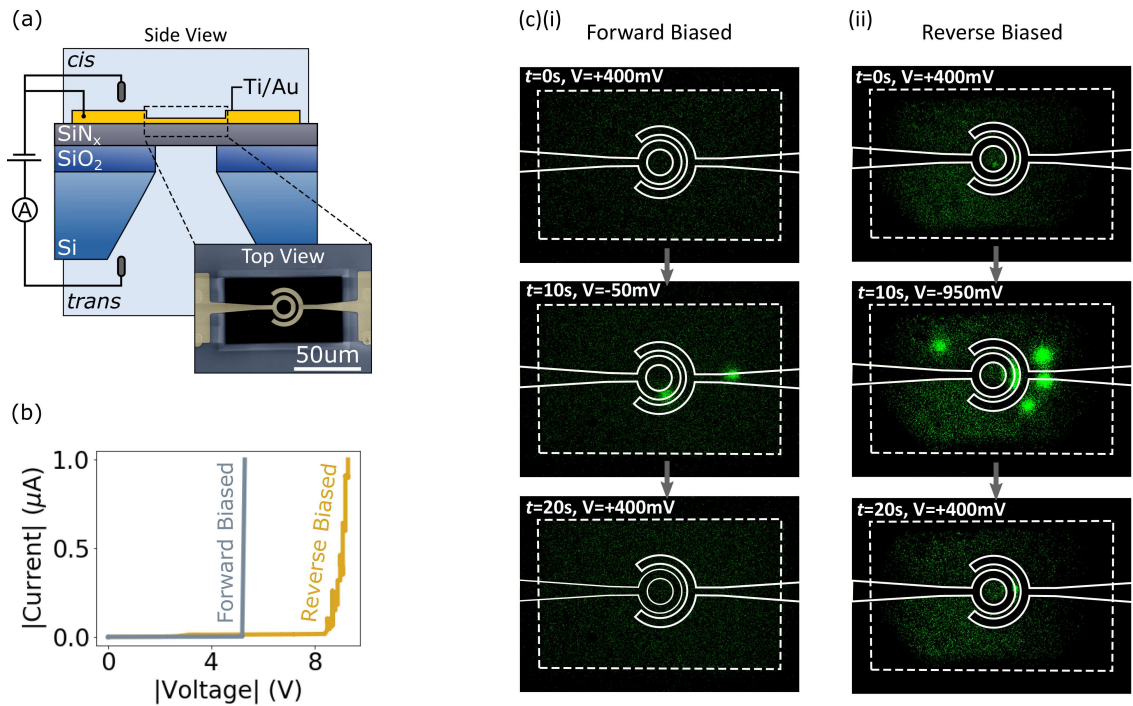}
    \caption{(a) Schematic of the experimental setup used for CBD on devices with microelectrodes on the membrane surface. The inset shows a false colour scanning electron micrograph of the electrode configuration over the suspended region of SiN$_x$. (b) Examples of CBD for devices with electrodes on the membrane surface when the electric field is applied in the forward-biased and the reverse-biased configuration. (c) Fluorescent microscopy images of the position of nanopores formed during CBD for the (i) forward-biased and (ii) reverse-biased configuration. The dashed white box shows the edges of the suspended region of SiN$_x$. The solid white lines depict the position of the electrodes. A time series of the images is shown for both breakdown conditions with a frame before, during, and after the application of a voltage that drives Ca$^{2+}$ ions through the nanopore.}
    \label{fmicroelec}
\end{figure*}

Another interesting observation from these measurements is that for the forward biased configuration, MIE devices breakdown at a much lower current density than MIM devices [Fig.~\ref{fmie}(a)(ii) and SI 5]. For instance, for the device shown in Fig.~\ref{fmie}(a)(ii), the MIE device undergoes breakdown at 8.1$\,$V (30$\,$nA) compared to 13$\,$V (613$\,$nA) for the MIM device. Although the reason for this is currently not clear, it suggests that for membranes exposed to an electrolyte solution, breakdown is not solely driven by the leakage current across the membrane as is typically assumed to be the case. Measuring conduction and breakdown in MIE devices as a function of electrolyte composition (e.g.$\,$pH) may provide further insight into the mechanism of nanopore formation during CBD. The use of MIE devices presented in this work would be useful in this regard as they enable isolation of the oxidation and reduction reactions.

\subsection{CBD with Microelectrodes on the Membrane Surface}

To further demonstrate how the oxidation reactions affect nanopore formation during CBD we have performed breakdown on devices with metal microelectrodes fabricated on the membrane surface. In contrast to the MIE devices considered in the previous section, a typical CBD configuration was used with electrolyte present on both sides of the membrane and the voltage applied via Ag/AgCl electrodes immersed in each reservoir. A schematic of the experimental setup is shown in Fig.~\ref{fmicroelec}(a). A false colour scanning electron micrograph of the electrode configuration over the suspended region of SiN$_x$ is shown in the inset of Fig.~\ref{fmicroelec}(a). To perform these experiments, the device was loaded into a fluidic cell with an integrated probe card that allows us to electrically contact each of the electrodes on the membrane surface. To avoid electrode delamination, the voltage of the on-chip electrodes and the voltage of the Ag/AgCl electrode in the \emph{cis} chamber are held at ground. The forward and reverse-biased configurations are then achieved by applying a positive or negative voltage respectively to the Ag/AgCl electrode in the
\emph{trans} chamber.

Figure \ref{fmicroelec}(b) shows typical conduction and breakdown events for devices in the forward and reverse-biased configuration. We observe that the device in the forward-biased configuration undergoes breakdown at a significantly lower voltage. This is consistent with the results shown in the previous section for the MIE devices. In particular, for the forward-biased configuration, electrons can be supplied to the SiN$_x$ from the electrodes on the membrane surface. As such, oxidation reactions do not need to occur at the membrane-electrolyte interface resulting in breakdown occurring at a lower voltage. Following this, one would expect nanopores to form only within the area covered by the electrodes for the forward-biased configuration.

To determine the position of the nanopores formed during CBD we have performed fluorescent microscopy to image the pores \cite{Anderson2014,Zrehen2017}. Here Ca$^{2+}$ ions are added to the solution on one side of the membrane while the Ca$^{2+}$ indicator dye Fluo-4 is added to the solution on the other side of the membrane. When a voltage is applied across the membrane, Ca$^{2+}$ ions can be driven through the pore resulting in a localised fluorescent signal at the nanopore. Figure \ref{fmicroelec}(c) shows fluorescent micrographs of the nanopores for the forward and reverse biased breakdown conditions. The white dashed box represents the edge of the suspended region of SiN$_x$ while the solid white lines represent the microelectrode edges. For each breakdown condition, three micrographs are shown representing a time series of data with a frame before, during, and after the application of a voltage that drives Ca$^{2+}$ ions through the nanopore. For the forward-biased configuration we observe that two nanopores form within the area covered by the electrodes on the membrane surface [Fig.~\ref{fmicroelec}(c)(i)]. Based on the area of the electrodes relative to the area of the membrane, the probability of this happening randomly is $\sim$1.6\%. However, for the reverse-biased configuration the nanopores form at random positions in the membrane [Fig.~\ref{fmicroelec}(c)(ii)]. As shown in SI 6, these results are reproducible across multiple devices. Note, for these experiments we intentionally did not reduce the voltage immediately after breakdown which resulted in the creation of multiple pores. This allowed us to obtain more statistics on the resulting nanopore position from a single membrane.

The formation of nanopores only within the area covered by the electrodes for the forward-biased configuration is consistent with the above results demonstrating that oxidation reactions at the membrane-electrolyte interface limit conduction during CBD. Namely, for the forward-biased configuration, electrodes on the membrane surface can supply electrons to the SiN$_x$. Therefore, an oxidation reaction does not need to occur resulting in breakdown occurring at a lower voltage in these regions. As a result, nanopores form only within the area covered by the electrodes on the membrane surface. For the reverse-biased configuration, an oxidation reaction must occur to inject electrons into the membrane (it is the reduction reaction that does not need to occur in the areas covered by the electrodes). As such, the nanopores form at random locations in the membrane.

These results are of practical importance for nanopore fabrication via CBD when micro/nanostructures are on the membrane surface. Nanopores integrated with complementary nanostructures have received interest in recent years as they have the ability to overcome issues associated with ionic current based nanopore sensing including limited device density \cite{Xie2011,Heerema2018} and low bandwidths \cite{Parkin2018,Bhat2018,Verschueren2018a}. Such complementary nanostructures include field-effect sensors \cite{Traversi2013,Xie2011,Graf2019,Fried2020}, tunnelling nanogaps \cite{Ivanov2011,Ivanov2014}, plasmonic nanostructures \cite{Spitzberg2019,Garoli2019}, radiofrequency antennas \cite{Bhat2018}, and dielectrophoretic electrodes \cite{Freedman2016}. To date, the development of these devices has been limited by the difficult fabrication processes that are required to integrate pores with complementary nanostructures \cite{Fried2018,Heerema2018,Healy2012}. Developing CBD techniques to self-align nanopores with complementary nanostructures is a promising way to overcome such issues \cite{Pud2015}. Our results demonstrate that nanopores can be localised to electrodes on the membrane simply by applying the appropriate polarity electric field. We also note that expansion of nanopores following CBD \cite{Beamish2012} is commonly performed using voltage pulses of alternating polarity \cite{Waugh2020,Leung2020}. The difference in the breakdown voltage depending on the electric field direction will therefore need to be taken into account when performing CBD on devices with electrodes on the membrane surface to avoid the unintentional formation of multiple pores.

\subsection{Varying the Membrane Stoichiometry}

We will now discuss how the stoichiometry of the SiN$_x$ membrane affects conduction and breakdown during CBD. Typically, nanopore experiments are performed using Si-rich SiN$_x$ membranes. This is due to the low intrinsic stress of these membranes which results in superior mechanical strength compared to stoichiometric Si$_3$N$_4$ membranes \cite{Habermehl1998, Edel2013}. That said, some nanopore studies have utilised stoichiometric SiN$_x$ membranes \cite{Li2001,Yanagi2018,Yanagi2019}. Other dielectrics such as HfO$_2$ are also becoming increasingly popular for solid-state nanopore membranes \cite{Larkin2013,Chou2020}. 

\begin{figure*}
    \includegraphics[width=16cm]{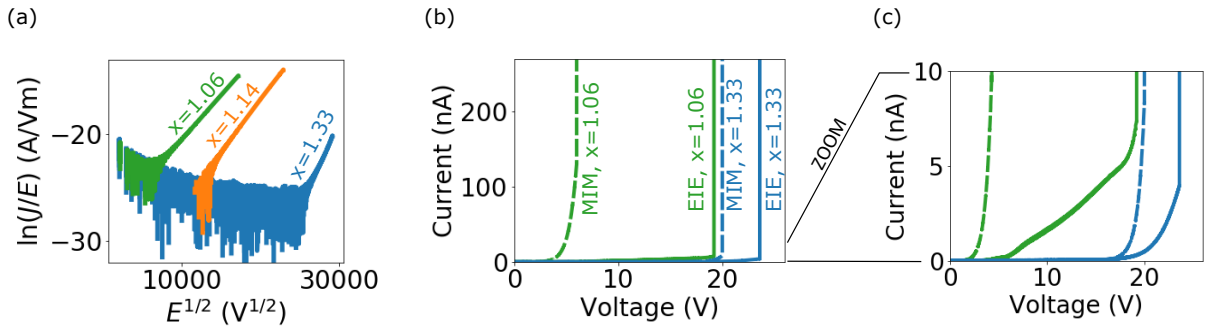}
    \caption{(a) PF plot of conduction in MIM devices for three different SiN$_x$ stoichiometries. (b) Comparision of the breakdown between MIM and EIE devices for two different SiN$_x$ stoichiometries. (c) Same data shown in (b) but with a reduced $y$-scale to enable visualisation of the conduction behaviour prior to breakdown in the EIE devices.}
    \label{fstoich}
\end{figure*}

Previous studies have demonstrated that electron transport through SiN$_x$ is significantly affected by the film stoichiometry \cite{Habermehl2002}. In particular, it has been shown that increasing the Si content of the SiN$_x$ film results in increased electron transport \cite{Habermehl2002}. This was thought to result from the decreased bond strain in Si-rich films which reduces the energy required to excite trapped electrons to the conduction band \cite{Habermehl2002}. This results in a lower electric field strength required to induce PF emission for Si-rich SiN$_x$. We have measured conduction in MIM devices for three different SiN$_x$ stoichiometries [Fig.~\ref{fstoich}(a)]. Consistent with previous studies \cite{Habermehl2002}, we observe an increase in conduction with increasing Si content.

We have also compared conduction and breakdown in EIE devices for SiN$_x$ membranes of different stoichiometries. Figure \ref{fstoich}(b) shows a comparison of conduction and breakdown in MIM and EIE devices for membranes with a N:Si ratio of $x$=1.06 (Si-rich) and $x$=1.33 (stoichiometric). As previously discussed, for Si-rich SiN$_x$ membranes the EIE device shows significantly less conduction and undergoes breakdown at a much larger voltage than the MIM device. However, this is no longer the case for stoichiometric Si$_3$N$_4$ membranes. Here, the MIM and EIE devices show comparable levels of conduction and breakdown occurs at a similar voltage. This results from the reduced electron transport through stoichiometric Si$_3$N$_4$, which now largely limits the electrical conduction in the EIE devices (rather than oxidation reactions at the membrane-electrolyte interface). 

The different conduction behaviour in EIE devices depending on the membrane stoichiometry is also highlighted in Fig.~\ref{fstoich}(c). This plot shows the same data as Fig.~\ref{fstoich}(b) with a reduced $y$-scale to enable visualisation of the conduction behaviour prior to breakdown in the EIE devices. For the Si-rich SiN$_x$, a leakage current can be measured at low voltages ($\sim$5$\,$V) since electron transport through these devices begins at low electric fields. The leakage current then increases slowly (approximately linearly) until breakdown due to the conduction being limited by oxidation reactions at the membrane interface. In contrast, for the stoichiometric membrane a leakage current can not be measured until large voltages are applied ($\sim$17$\,$V). This is because of the reduced electron transport in this dielectric which results in negligible conduction up to large electric fields. After the onset of conduction, the current increases approximately exponentially until breakdown (albeit at a slower rate than the MIM device since oxidation reactions still limit the current to some extent). Similar stoichiometry dependent conduction characteristics have been observed consistently for many devices (SI 7). These results highlight the material dependent nature of the conduction and nanopore formation processes during CBD. As such, these results will be of interest as this technique is further developed to fabricate nanopores in previously unexplored material systems. 

\section{Conclusion}

To understand the process of nanopore formation during CBD we have studied conduction and breakdown in SiN$_x$ membranes when the voltage is applied via (i) metal electrodes on the membrane surface, (ii) electrolyte solutions, and (iii) a combination of the two. We demonstrate that, for Si-rich SiN$_x$ membranes, oxidation reactions limit the electrical conduction across the membrane during CBD. As a result, when performing CBD with electrodes on the membrane surface we can remove the need for oxidation reactions (since electrons can be supplied by the metal) enabling nanopore formation to be localised to the area covered by the electrodes. We also studied conduction and breakdown when varying the stoichiometry of the SiN$_x$ membrane. Here, we show that stoichiometric Si$_3$N$_4$ displays significantly decreased electron transport across the dielectric compared to Si-rich SiN$_x$. As a result, it is electron transport across the dielectric which largely limits the electrical conduction in these membranes (rather than oxidation reactions at the membrane-electrolyte interface). This demonstrates the highly material dependent nature of conduction and nanopore formation during CBD. Our results are important in further understanding the mechanism by which nanopores are formed during CBD which will be necessary to further develop this technique in the coming years. For instance, understanding our results will be crucial in developing CBD techniques to create nanopores integrated with complementary nanostructures on the membrane surface. Our results will also be of interest to researchers aiming to develop reliable CBD techniques for different membrane materials. 

\section{Methods}

\subsection{Device Fabrication}

Devices were fabricated on double-side polished Si wafers with a crystal orientation of $<$100$>$ and resistivity of 1-100$\,\Omega$.cm. A wet thermal oxide layer of thickness 500$\,$nm is grown on both sides of the wafer. Low pressure chemical vapour deposition (LPCVD) is used to deposit a 20-25$\,$nm thick SiN$_x$ layer on both sides of the wafer. The stoichiometry of the SiN$_x$ film was varied by controlling the ratio of SiCl$_2$H$_2$ and NH$_3$ during deposition. Photolithography and reactive ion etching to remove the SiN$_x$ and SiO$_2$ from the backside of the wafer is used to create a hard mask. A polymethylmethacrylate (PMMA) layer is spun on top of the SiN$_x$ to protect the film during subsequent etching steps. Anisotropic etching of Si followed by isotropic etching of the SiO$_2$ in 30\% KOH at 80$^{\circ}$C then creates suspended SiN$_x$ membranes. This is the final device geometry use for the EIE devices. MIM and MIE devices are created from this device geometry by depositing a 5/45$\,$nm Cr/Au metal layer on one or both sides of the membrane via thermal evaporation. The edges of the device were covered with polyimide (Kapton) tape during evaporation to avoid shorting of the electrodes. 

Devices with microelectrodes on the membrane surface were fabricated via a similar process to that described above. However, here, after SiN$_x$ deposition metal electrodes are deposited on the front of the wafer via electron beam lithography (EBL) and electron beam evaporation followed by photolithography and electron beam evaporation. The thickness of the electrodes are 5/95$\,$nm Ti/Au for the regions defined by photolithography and 5/15$\,$nm Ti/Au for the regions defined by EBL. The rest of the process then proceeds as described above.

\subsection{Conduction and breakdown measurements}

For measurements using the EIE configuration, devices were first cleaned in Piranha solution (ratio of 3:1 H$_2$SO$_4$:H$_2$O$_2$). Devices were then loaded into a fluidic cell (purchased from Nanopore Solutions) and each reservoir filled with a 1$\,$M KCl, 10mM$\,$Tris, 0.1$\,$mM EDTA buffer solution at pH 8. Ag/AgCl electrodes were then inserted into each reservoir. For measurements on the MIM devices, one side of the device was adhered to a contact pad on a printed circuit board (PCB) using silver paste. The electrode on the other side of the device was then wirebonded to another contact pad on the PCB. For measurements on the MIE devices, the metal electrode was wirebonded to a contact pad of a PCB. The PCB was then loaded into a custom made fluidic cell and the reservoir filled with 1$\,$M KCl, 10$\,$mM Tris, 0.1$\,$mM EDTA buffer at pH 8. A Ag/AgCl electrode was then inserted into this reservoir. For CBD measurements on devices with microelectrodes on the membrane surface devices were loaded into a fluidic cell with an integrated probe card that contacts each of the electrodes (designed in collaboration with Nanopore Solutions). Reservoirs on both side of the membrane are then filled with 1$\,$M KCl, 10$\,$mM Tris, 0.1$\,$mM EDTA buffer at pH 8. Ag/AgCl electrodes were then inserted into each reservoir. The same protocol was used to measure conduction and breakdown in all device geometries. Namely, a voltage ramp increasing in steps of 100$\,$mV every 4$\,$s is applied across the membrane while simultaneously measuring the current using a Keithley 2450 source metre.

\subsection{Fluorescent Imaging of Nanopores}

To perform fluorescent imaging of devices with microelectrodes on the membrane surface after CBD, devices were cleaned in DI water, followed by acetone, and O$_2$ plasma etching. Prior to measurements the devices were again cleaned using UV-ozone treatment. The devices were then adhered onto a custom-built device holder. The device holder was in turn mounted onto an inverted microscope (IX71, Olympus, USA). For fluorescence imaging, the device was illuminated by with a fibre-coupled 488$\,$nm tunable Argon ion laser (Model 35-LAP-431-230, Melles Griot). A 498$\,$nm dichroic mirror reflected the incoming light towards the sample, where a 60x objective (UPLSAPO 60XW, Olympus, USA) was employed to both illuminate the sample and to collect the emitted fluorescence. The $cis$ and $trans$ chambers were filled with CaCl$_2$ solution (50$\,\mu$M CaCl$_2$, 100$\,$mM KCl in DI water) and Fluo-4 solution (5$\,\mu$M Fluo-4, 100$\,$mM KCl in DI water) respectively. Ag/AgCl electrodes were inserted into both chambers and connected to an eONE (Elements srl) current amplifier. A negative voltage was applied to the Ag/AgCl electrode in the $trans$ chamber to electrophoretically drive Ca$^{2+}$ through the nanopore. Transport of Ca$^{2+}$ ions from the $cis$ chamber to the $trans$ chamber activates the Ca$^{2+}$ dependent Fluro-4 resulting in a highly localised and voltage-tunable fluorescent spot at the nanopore which was recorded by an electron multiplying charge coupled device camera (Photometrics Cascade II, USA).

\section{Conflicts of Interest}

J. Y. is a principal in Nanopore Solutions whose fluidic devices were used in this study. All other authors have no conflicts to declare. 

\section{Acknowledegments}

Substrate, membrane and some of the electrode fabrication was conducted at the Center for Nanophase Materials Sciences, which is a DOE Office of Science User Facility. J. F. thanks the Oxford Australia Scholarship committee and the University of Western Australia for Funding. J. Y. was funded by an FCT contract according to DL57/2016, [SFRH/BPD/80071/2011]. Work in J.Y.’s lab  was funded by national funds through FCT - Fundação para a Ciência e a Tecnologia, I. P., Project MOSTMICRO-ITQB with refs UIDB/04612/2020 and UIDP/04612/2020 and Project PTDC/NAN-MAT/31100/2017. J. M. was supported through the UKRI Future Leaders Fellowship, Grant No. MR/S032541/1, with in-kind support from the Royal Academy of Engineering. A. I. and J. E. acknowledge support from BBSRC grant BB/R022429/1, EPSCR grant EP/P011985/1, and Analytical Chemistry Trust Fund grant 600322/05. This project has also received funding from the European Research Council (ERC) under the European Union's Horizon 2020 research and innovation programme (grant agreement No 724300 and 875525).

\bibliography{refs}

\end{document}